# Growing fluctuation of quantum weak invariant and dissipation

Sumiyoshi Abe [a-d]

[a] *Department of Physics, College of Information Science and Engineering, Huaqiao University, Xiamen 361021, China*
[b] *Institute of Physics, Kazan Federal University, Kazan 420008, Russia*
[c] *Department of Natural and Mathematical Sciences, Turin Polytechnic University in Tashkent, Tashkent 100095, Uzbekistan*
[d] *ESIEA, 9 Rue Vesale, Paris 75005, France*

**Abstract**   The concept of weak invariants has recently been introduced in the context of conserved quantities in finite-time processes in nonequilibrium quantum thermodynamics. A weak invariant itself has a time-dependent spectrum, but its expectation value remains constant under time evolution defined by a relevant master equation. Although its expectation value is thus conserved by definition, its fluctuation is not. Here, time evolution of such a fluctuation is studied. It is shown that if the subdynamics is given by a completely positive map, then the fluctuation of the associated weak invariant does not decrease in time. It is also shown, in the case of the Lindblad equation, how the growth rate of the fluctuation is connected to the dissipator. As examples, the harmonic oscillator with a time-dependent frequency and the spin in a varying external magnetic field are discussed, and the fluctuations of their Hamiltonians as the weak invariants are analyzed. Furthermore, a general relation is presented for the specific heat and temperature of any subsystem near equilibrium following the slow Markovian isoenergetic process.

# 1. Introduction

Although studies of open quantum systems have a long tradition [1,2], they seem to be at the stage of new importance today. An integral part of the reasons behind this fact may be due to rapidly-developing quantum thermodynamics [3-5] and its fundamental relevance to quantum information and quantum computation [6,7]. For any quantum information processing in reality can hardly be taken place freely from environmental effects. A classical system treated by equilibrium thermodynamics is described by a set of dynamical/thermodynamical variables. In quantum thermodynamics, on the other hand, not only dynamical variables but also the Hilbert space are needed. This "double structure" can lead to diverse concepts of baths. Examples often discussed in the literature are the dephasing bath [8] and the energy bath [9] that may not naively have their classical counterparts. The role of the dephasing bath is to realize decoherence of a quantum state. In this respect, we wish to point out that there exists the notion of dissipationless decoherence [10]. The energy bath exchanges with a subsystem the energy *not purely in the form of heat*. Recent developments in energy transfer technique [11,12] should offer a physical basis to such a bath. Thus, such a bath plays an active role, in contrast to the passive role of the heat bath in classical thermodynamics.

Thermodynamic processes are characterized by conservations of quantities or variables, such as in the isothermal, isoentropic, isoenergetic, isochoric, and isobaric processes. Among these, the isoenergetic process stands out, since it is relevant to the energy bath mentioned above. The expectation value of the subsystem Hamiltonian as the internal energy is kept constant along this process. It is noted that, in quantum thermodynamics, it is different from the isothermal process because of the quantum-mechanical violation of the law of equipartition of energy. Expansion of size of the subsystem shifts the spectrum of the Hamiltonian lower, but energy transfer from the energy bath can compensate it, for example. As a result, the isoenergetic process highlights in a peculiar manner how quantum thermodynamics widens the view of traditional concepts in classical thermodynamics. For example, a quantum engine with three processes including the isoenergetic process may possess some specific properties in its performance [13]: the efficiency becomes maximum if the expansion ratio of the engine is appropriately tuned, and moreover the lower the temperature is, the higher the maximum efficiency becomes. These are exotic in view of classical thermodynamics.

The concept of quantum weak invariants has been introduced under this background in quantum thermodynamics [14].

A time-dependent observable is said to be a strong invariant if its spectrum does not depend on time. A celebrated example is the Lewis-Riesenfeld invariant [15] associated with the unitary dynamics (see also Ref. [16] for extensive discussions about strong invariants in nonstationary systems). It is known [17] that the strong invariant of the time-dependent harmonic oscillator can be obtained through Noether's theorem (see also Ref. [18] for a simplified discussion). The Lewis-Riesenfeld invariant has been applied to a variety of problems including geometric phases [19-21], coherent and squeezed states [22-25], Grassmann fermions [26], quantum computation [27], cooling atoms in harmonic traps [28], population inversion [29], nonstationary quantum field theory in electromagnetic and cosmological backgrounds [30-32], and third quantized quantum cosmology [33].

On the other hand, a time-dependent observable is referred to as a weak invariant if its spectrum varies in time but its expectation value is conserved. The study of weak invariants is still at its early stage compared to the diversity of works done on strong invariants. However, the results presented in Refs. [34,35] show that a weak invariant can be interpreted as the Noether charge in the action principle for kinetic theory, suggesting possible existence of some rich structure to be revealed.

The purpose of the present work is as follows. The expectation value of a weak invariant is kept constant in time by definition, but the fluctuation may not. Here, we discuss time evolution of the fluctuation of a quantum weak invariant. This issue may particularly be relevant to quantum thermodynamics of small systems, in which the effects of fluctuations are nonnegligible, in general. We prove that the fluctuation of a quantum weak invariant monotonically grows in time if the subdynamics is given by a completely positive map. As an important case, we consider the Lindblad equation [36,37] and show how the growth rate is related to the dissipator contained in the equation. Then, we discuss the harmonic oscillator with a time-dependent frequency and the spin in a varying magnetic field and evaluate the growth rates of the fluctuations of their Hamiltonians as the weak invariants. Finally, we present a general relation to be satisfied by the changes of the slowly-time-dependent specific heat and temperature of any subsystem in the near-equilibrium state following the Markovian isoenergetic process.

## 2. Quantum weak invariant

A quantum weak invariant, $I(t)$, is a Hermitian operator and is a function of relevant observables as dynamical variables. It is not an invariant operator, but its expectation value, $\langle I(t) \rangle = \text{tr}\big(I(t)\rho(t)\big)$, is conserved [14]:

$$\frac{d\langle I(t) \rangle}{dt} = 0, \qquad (1)$$

where $\rho(t)$ is a density matrix describing a quantum state of a subsystem, which is positive semidefinite, normalized and obeys a certain time-evolution equation. Although $I(t)\rho(t)$ has to be trace-class, $I(t)$ itself is not necessarily so.

First, let us consider this concept in a more general framework. We require finite-time evolution of the density matrix to be described by a completely positive map, $\Phi_{t',t}$, in the Kraus representation [38]

$$\Phi_{t',t}: \rho(t) \to \rho(t') = \Phi_{t',t}\big(\rho(t)\big) = \sum_k V_k(t',t)\,\rho(t)\,V_k^\dagger(t',t) \qquad (2)$$

with the trace-preserving condition

$$\sum_k V_k^\dagger(t',t)\,V_k(t',t) = \mathbb{I}, \qquad (3)$$

where $t' > t$ and $\mathbb{I}$ stands for the identity operator. Equation (3) implies that the set $\{V_k^\dagger(t',t) V_k(t',t)\}_k$ forms a positive operator-valued measure. Using this map, we describe the weak invariant as follows:

$$\mathrm{tr}\big(I(t')\rho(t')\big) = \mathrm{tr}\big(I(t)\rho(t)\big). \tag{4}$$

The left-hand side of this equation is rewritten as

$$\mathrm{tr}\big(I(t')\rho(t')\big) = \mathrm{tr}\big\{I(t')\, \Phi_{t',t}\big(\rho(t)\big)\big\} = \mathrm{tr}\big\{\Phi_{t',t}^*\big(I(t')\big)\rho(t)\big\}, \tag{5}$$

where $\Phi_{t',t}^*$ denotes the adjoint of $\Phi_{t',t}$ in Eq. (2), that is,

$$\Phi_{t',t}^*\big(I(t)\big) = \sum_k V_k^\dagger(t',t)\, I(t)\, V_k(t',t). \tag{6}$$

Therefore, $I(t)$ is a weak invariant associated with the map, $\Phi_{t',t}$, if it satisfies the following equation:

$$\Phi_{t',t}^*\big(I(t')\big) = I(t). \tag{7}$$

A crucial point of importance for the later discussion is that, because of Eq. (3), the identity operator is a fixed point of the adjoint map, $\Phi_{t',t}^*$, that is,

$$\Phi^*_{t',t}(\mathbb{I}) = \mathbb{I}. \tag{8}$$

Therefore, $\Phi^*_{t',t}$ is not only completely positive itself but also unital [39]. Equation (7) combined with these properties of the adjoint map provides the concept of quantum weak invariants with a new basis.

Next, let us derive the Lindblad equation from Eq. (2). Although such an issue is known in the literature, here we do so in order to view the above discussion from another perspective. Consider infinitesimal time evolution: $t' = t + \Delta t$. Pick up one of $V_k$'s, say $V_0$, and write it and the rest respectively as follows [40]:

$$V_0(t+\Delta t, t) = \mathbb{I} - i\Delta t\, H(t) - \frac{\Delta t}{2} \sum_{k \neq 0} |g_k|^2 L_k^\dagger L_k + O\big((\Delta t)^2\big), \tag{9}$$

$$V_n(t+\Delta t, t) = (\Delta t)^{1/2} g_n L_n + O(\Delta t) \qquad (k = n \neq 0). \tag{10}$$

where $g_k$'s are *c*-number coefficients, and here and hereafter, $\hbar$ is set equal to unity. Like the subsystem Hamiltonian, $H(t)$, the coefficients as well as the operators, $L_k$'s, may explicitly depend on time. Locality in time implies that we are working in the Markovian approximation. Substituting Eqs. (9) and (10) into Eq. (2), we have

$$\rho(t+\Delta t) = \rho(t) - i\Delta t \big[H(t), \rho(t)\big]$$
$$-\Delta t \sum_n c_n \Big(L_n^\dagger L_n \rho(t) + \rho(t) L_n^\dagger L_n - 2 L_n \rho(t) L_n^\dagger\Big) + O\big((\Delta t)^{3/2}\big), \tag{11}$$

where $c_n$'s are nonnegative coefficients given by $c_n = |g_n|^2 / 2$. Then, in the limit $\Delta t \to 0+$, we obtain from Eq. (11) the Lindblad equation [36,37]

$$i \frac{\partial \rho(t)}{\partial t} = [H(t), \rho(t)] - i \sum_n c_n \left( L_n^\dagger L_n \rho(t) + \rho(t) L_n^\dagger L_n - 2 L_n \rho(t) L_n^\dagger \right), \quad (12)$$

which is known to be the most general linear Markovian quantum master equation. Nonnegativity of $c_n$'s is essential for the equation to preserve positive semidefiniteness of the density matrix. The second term on the right-hand side is refereed to as the dissipator and is responsible for the nonunitary subdynamics. To determine the operators, $L_n$'s, termed the Lindbladian operators, as well as the *c*-number coefficients, it is necessary to have detailed information about an interaction between the subsystem and the environment, in general. However, under a certain circumstance, they can uniquely be determined without such information. This point will be discussed in Sec. IV.

Following a similar procedure, we obtain from Eqs. (6) and (7) the following equation for the weak invariant:

$$\frac{\partial I(t)}{\partial t} + i [H(t), I(t)] - \sum_n c_n \left( L_n^\dagger L_n I(t) + I(t) L_n^\dagger L_n - 2 L_n^\dagger I(t) L_n \right) = 0. \quad (13)$$

Contrarily, we can ascertain Eq. (1) using Eqs. (12) and (13).

Three comments on Eq. (13) are in order. This equation has actually been derived in Ref. [41] in a different way in order to make a basis as the eigenstates of $I(t)$ to write down the elements of a density matrix. Secondly, Eq. (13) possesses symmetry. Under the shift,

$$I(t) \rightarrow I(t) + a \mathbb{I} \tag{14}$$

with $a$ being constant, Eq. (13) remains unchanged. Thirdly, if the weak invariant and all of the Lindbladian operators commute with each other, then the third term on the left-hand side vanishes, and accordingly Eq. (13) becomes reduced to $\frac{\partial I(t)}{\partial t} + i[H(t), I(t)] = 0$, which is the equation for the Lewis-Riesenfeld strong invariant [15]. In this case, the formal solution satisfying an appropriate initial condition, $I(0)$, is given by $I(t) = U(t) I(0) U^{\dagger}(t)$, where $U(t)$ is the unitary time evolution operator given by $U(t) = T \exp\left(-i \int_0^t ds\, H(s)\right)$, where $T$ is the chronological symbol for time-ordered products. Therefore, the spectrum of the invariant remains unchanged under time evolution. In what follows, such a case will be excluded.

3. **Monotonically growing fluctuation of weak invariant and dissipation**

Now, let us discuss time evolution of the fluctuation of the weak invariant, $I(t)$. Here, we employ the variance

$$(\Delta I)^2 (t) = \langle I^2(t) \rangle - \langle I(t) \rangle^2 \tag{15}$$

in order to quantify the fluctuation. The second term on the right-hand side is constant in time, by definition.

In parallel with the discussion in the previous section, first, let us consider the maps in Eqs. (2) and (6). As mentioned above, what to be examined is the second moment

$$\left\langle I^2(t') \right\rangle = \text{tr}\left(I^2(t') \rho(t')\right) = \text{tr}\left\{I^2(t') \Phi_{t',t}\left(\rho(t)\right)\right\} = \text{tr}\left\{\Phi^*_{t',t}\left(I^2(t')\right) \rho(t)\right\}. \quad (16)$$

From Kadison's theorem [39] for a completely positive unital map and operator convex $I^2(t)$ with $I(t)$ being Hermitian, it follows that

$$\Phi^*_{t',t}\left(I^2(t')\right) \geq \left[\Phi^*_{t',t}\left(I(t')\right)\right]^2 = I^2(t), \quad (17)$$

where Eq. (7) has been used in the last equality. (The operator inequality $A \geq B$ implies that both $A$ and $B$ are Hermitian and all eigenvalues of $A - B$ are nonnegative.) Using this in Eq. (16), we have $\left\langle I^2(t') \right\rangle \geq \left\langle I^2(t) \right\rangle$. Therefore, we obtain the following general result:

$$\left(\Delta I\right)^2(t') \geq \left(\Delta I\right)^2(t) \qquad (t' > t), \quad (18)$$

showing that the fluctuation of the weak invariant never decreases. The equality holds for the unitary case corresponding to the strong invariant and thus can actually be excluded.

Next, let us evaluate the growth rate of the fluctuation. As repeatedly mentioned, $\langle I(t) \rangle$ does not depend on time. Using Eqs. (12) and (13), we obtain another result:

$$\frac{d(\Delta I)^2(t)}{dt} = \frac{d\langle I^2(t) \rangle}{dt} = 2 \sum_n c_n \langle [L_n, I(t)]^\dagger [L_n, I(t)] \rangle. \quad (19)$$

This highlights how the growth rate is always positive and is connected to the dissipator. It is noted that the Hamiltonian does not directly contribute to it. This fact might be somewhat analogous to diffusion, to which mechanical forces do not usually contribute. As found in Ref. [35], the weak invariant (in classical kinetic theory) can be regarded as the Noether charge. In this respect, the commutator, $[L_n, I]$, appearing in Eq. (19) is quantum-mechanically interpreted as a deformation of the Lindbladian operators through a gauge transformation generated by the Noether charge.

It is of interest to compare Eq. (19) with the entropy production rate. In Ref. [42], it is proved that, under the Lindblad equation in Eq. (12), the von Neumann entropy, $S[\rho(t)] = -\mathrm{tr}(\rho(t) \ln \rho(t))$, evolves as

$$\frac{dS[\rho(t)]}{dt} \geq 2 \sum_n c_n \langle [L_n^\dagger, L_n] \rangle. \quad (20)$$

This relation has recently been generalized to the case of the Rényi entropy, $S_\alpha[\rho(t)] = (1-\alpha)^{-1} \ln(\mathrm{tr}\, \rho^\alpha(t))$ ($\alpha > 0$), as follows [43]:

$$\frac{d S_\alpha[\rho(t)]}{dt} \geq 2 \sum_n c_n \left\langle \left[ L_n^\dagger, L_n \right] \right\rangle_\alpha, \qquad (21)$$

where the symbol appearing on the right-hand side is the so-called $\alpha$-expectation value, $\left\langle * \right\rangle_\alpha = \mathrm{tr}\left( * P^{(\alpha)}(t) \right)$, with $P^{(\alpha)}(t)$ being the *escort density matrix* defined by $P^{(\alpha)}(t) \equiv \rho^\alpha(t) / \mathrm{tr}\, \rho^\alpha(t)$. Equations (20) and (21) are inequalities, whereas Eq. (19) is an equality. If the subdynamics is unital, then the Lindbladian operators are normal, i.e., $\left[ L_n^\dagger, L_n \right] = 0$, and accordingly the entropy does not decrease. However, if the subdynamics is not unital, then the lower bounds in Eqs. (20) and (21) can be negative, in general. On the other hand, the growth rate in Eq. (19) is always positive even for the nonunital subdynamics. These should be regarded as the salient features of the fluctuation of the weak invariant.

## 4. Time-dependent Hamiltonian as weak invariant: Examples of harmonic oscillator and spin

Weak invariants have been applied to isoenergetic processes in finite-time quantum thermodynamics. The isoenergetic process is the one, along which the internal energy, $U = \left\langle H(t) \right\rangle$, is kept constant. Therefore, the time-dependent subsystem Hamiltonian is a weak invariant. Examples treated here are the harmonic oscillator with a time-dependent frequency [44] and the Pauli spin in a varying external magnetic field [45]. Physically, the former may be concerned with a small engine made of an ion in a time-dependent harmonic trap, whereas the latter is related actually not only to the spin but also to a two-level atom, both of which can experimentally be used for realizations of quantum heat engine [46,47]. The finite power outputs and the works done along the isoenergetic processes have been calculated in Refs. [44,45] with use of the Lindblad equations in the temporally-local equilibrium approximation. However, as mentioned in Sec. 1, the fluctuations of quantities are important in small systems. In this section, we discuss the fluctuations of the time-dependent subsystem Hamiltonians as the weak invariants and their growth rates.

Let us impose the condition that the time-dependent subsystem Hamiltonian, $H(t)$, is a weak invariant associated with the Lindblad equation. Then, from Eq. (13), it follows that

$$\frac{\partial H(t)}{\partial t} - \sum_n c_n \left( L_n^\dagger L_n H(t) + H(t) L_n^\dagger L_n - 2 L_n^\dagger H(t) L_n \right) = 0 . \qquad (22)$$

This equation has been studied in Ref. [48]. An observation made there is that Eq. (22) acquires a peculiar mathematical property when the Lindbladian operators are Hermitian (i.e., the subdynamics is unital) and there exists a spectrum generating algebra, e.g., the subsystem Hamiltonian is Lie-algebra-valued. This is precisely the case in the harmonic oscillator with a time-dependent frequency and the Pauli spin in a varying external magnetic field.

If the Lindbladian operators are Hermitian, then they are normal, and the entropy production rates in Eqs. (20) and (21) are nonnegative. And, Eq. (22) becomes

$$\frac{\partial H(t)}{\partial t} - \sum_n c_n \left[ L_n, \left[ L_n, H(t) \right] \right] = 0. \qquad (23)$$

Dissipationlss decoherence is the peculiar case when the Lindbladian is the subsystem Hamiltonian itself [10], and therefore Eq. (23) forces it to be time-independent. In other words, as long as the subsystem Hamiltonian is a time-dependent weak invariant, decoherence induces dissipation. Also, it has been discussed in Ref. [48] how Eq. (23) can determine the Lindbladian operators and the *c*-number coefficients without detailed information about the interaction between the subsystem and the environment if the subsystem Hamiltonian is Lie-algebra-valued.

Thus, the first example is the time-dependent harmonic oscillator with unit mass. Its subsystem Hamiltonian reads

$$H(t) = K_1 + k(t) K_2. \qquad (24)$$

$k(t)$ is related to the frequency $\omega(t)$ as $k(t) = \omega^2(t)$. $K_1$ and $K_2$ are given in terms of the momentum operator, *p*, and position operator, *x*, as $K_1 = p^2/2$ and $K_2 = x^2/2$. Together with $K_3 = (px + xp)/2$, they form a closed algebra: $[K_1, K_2] = -iK_3$, $[K_2, K_3] = 2iK_2$, $[K_3, K_1] = 2iK_1$. Formally, this is isomorphic to the $su(1,1)$ Lie algebra [49]. In this case, there exists a single Lindbladian operator, which is $L = K_2$ and the corresponding *c*-number coefficient is $c = -\dot{k}(t)/2$, where the over-dot denotes the time derivative. Since *c* has to be positive, the frequency has to decrease monotonically in time

$$\dot{k}(t) < 0, \tag{25}$$

implying that the potential of the harmonic trap, for example, has to be widening. Since $[L, H(t)] = [K_2, K_1 + k(t) K_2] = i K_3$, it follows from the general formula in Eq. (19) that

$$\frac{d(\Delta H)^2}{dt} = -\dot{k}(t) \langle K_3^2 \rangle, \tag{26}$$

which is positive because of Eq. (25).

The second example, which is the Pauli spin in a varying external magnetic field, is described by the subsystem Hamiltonian

$$H(t) = \mathbf{B}(t) \cdot \sigma, \tag{27}$$

where $\sigma = (\sigma_1, \sigma_2, \sigma_3)$ is the Pauli-matrix vector and $\mathbf{B}(t)$ is the magnetic field. The constant factor involving the gyromagnetic ratio is set equal to unity for simplicity. The spectrum generating algebra is $su(2)$. The corresponding Lindbladian operators are: $L_n = \sigma_n$ ($n = 1, 2, 3$), which are Hermitian. Therefore, from Eqs. (23) and (27), we have $\dot{\mathbf{B}}(t) = 4\big((c_2 + c_3) B_1, (c_3 + c_1) B_2, (c_1 + c_2) B_3\big)$, from which we further obtain the following three time-dependent $c$-number coefficients in the Lindblad equation:

$$c_1 = \frac{1}{8}\left(-\frac{\dot{B}_1}{B_1} + \frac{\dot{B}_2}{B_2} + \frac{\dot{B}_3}{B_3}\right), \quad c_2 = \frac{1}{8}\left(\frac{\dot{B}_1}{B_1} - \frac{\dot{B}_2}{B_2} + \frac{\dot{B}_3}{B_3}\right), \quad c_3 = \frac{1}{8}\left(\frac{\dot{B}_1}{B_1} + \frac{\dot{B}_2}{B_2} - \frac{\dot{B}_3}{B_3}\right). \quad (28)$$

On the other hand, a straightforward calculation gives $[L_n, H(t)] = 2i(\mathbf{B}(t) \times \sigma)_n$, which leads to $[L_n, H(t)]^\dagger [L_n, H(t)] = 4(\delta_{nn}\mathbf{B}^2 - B_n B_n)\mathbb{I}$. Therefore, Eq. (19) gives rise to the result:

$$\frac{d(\Delta H)^2(t)}{dt} = \frac{d\mathbf{B}^2(t)}{dt}, \quad (29)$$

which is, in fact, positive, since $c_n$'s in Eq. (28) are nonnegative and not all of them can vanish. In contrast to the oscillator case in Eq. (26), Eq. (29) does not have state dependence. Positivity of the right-hand side means that it is possible for the isoenergetic process to be realized if and only if the strength of the magnetic field monotonically increases in time.

### 5. Relation between specific heat and temperature in slow Markovian isoenergetic process

An immediate outcome from the result in Eq. (19) is the following. In canonical ensemble of the strict equilibrium states with no explicit time dependence, the fluctuation of the subsystem Hamiltonian is related to the specific heat (at constant volume), $C$, as $(\Delta H)^2 = T^2 C$, where $T$ is temperature and the Boltzmann constant is set equal to unity. In the slowly-time-dependent states slightly out of equilibrium, description of the subsystem by canonical ensemble is still good in the (temporally-)local equilibrium approximation. Thus, both *T* and *C* slowly vary in time. Applying the result in Eq. (19) to the time-dependent subsystem Hamiltonian as a weak invariant, we find

$$2C\frac{dT}{dt} + T\frac{dC}{dt} > 0, \tag{30}$$

which has to hold along any slow Markovian isoenergetic process followed by the subsystem in a near-equilibrium state.

## 6. Conclusion

We have shown that the fluctuation of the weak invariant associated with the subdynamics described by a completely positive map monotonically grows, in general. We have found in the case of the Lindblad equation how the growth rate of the fluctuation is connected to dissipation. We have discussed a peculiar circumstance, under which the Lindbladian operators are Hermitian and the time-dependent Hamiltonian as a weak invariant is Lie-algebra valued. We have also presented a relation to be satisfied by the specific heat and temperature of a subsystem in a state near equilibrium that follow any slow Markovian isoenergetic process.

**Acknowledgments**

This work has been supported in part by a grant from National Natural Science Foundation of China (No. 11775084), the Program of Fujian Province, and the Program of Competitive Growth of Kazan Federal University from the Ministry of Education and Science of the Russian Federation, to all of which the author gratefully acknowledges.

**Appendix**

In Sec. I, the "double structure" in quantum thermodynamics is pointed out. It seems worth noting that a situation somewhat analogous to this actually appears also in classical thermodynamics in the nonequilibrium regime since the kinetic approach contains both variables and distributions [35,50]. Here, let us briefly look at a classical weak invariant associated with the Fokker-Planck equation [51]:

$$\frac{\partial P(\mathbf{x},t)}{\partial t} = -\sum_{i=1}^{N} \frac{\partial}{\partial x_i}\left(K_i(\mathbf{x},t) P(\mathbf{x},t)\right) + \sum_{i,j=1}^{N} \frac{\partial^2}{\partial x_i \partial x_j}\left(D_{ij}(\mathbf{x},t) P(\mathbf{x},t)\right), \quad \text{(A.1)}$$

where $P(\mathbf{x},t) d^N \mathbf{x}$ is the probability of finding the $N$-tuple of the variables $\mathbf{x} = (x_1, x_2, ..., x_N)$ in the region $[x_1, x_1 + dx_1] \times [x_2, x_2 + dx_2] \times \cdots \times [x_N, x_N + dx_N]$ at time $t$. On the right-hand side, the first one is the drift term, and $D = (D_{ij})$ in the second one is a symmetric and positive definite diffusion matrix. A weak invariant associated with this kinetic equation is a quantity, $J(\mathbf{x},t)$, satisfying

$$\frac{\partial J(\mathbf{x},t)}{\partial t} + \sum_{i=1}^{K} K_i(\mathbf{x},t) \frac{\partial J(\mathbf{x},t)}{\partial x_i} + \sum_{i,j=1}^{N} D_{ij}(\mathbf{x},t) \frac{\partial^2 J(\mathbf{x},t)}{\partial x_i \partial x_j} = 0. \qquad (A.2)$$

Under the assumption that the probability distribution and its derivatives vanish sufficiently rapid in the limit $|\mathbf{x}| \to \infty$, the expectation value of the weak invariant,

$$\overline{J}(t) = \int d^N\mathbf{x}\, J(\mathbf{x},t)\, P(\mathbf{x},t) \qquad (A.3)$$

with the domain of integration being the whole *N*-dimensional space, is seen to be conserved in time:

$$\frac{d\overline{J}(t)}{dt} = 0. \qquad (A.4)$$

Boundary conditions do not have to be imposed on the weak invariant, as long as the probability distribution vanishes sufficiently rapid at the spatial infinity. Therefore, there exists freedom in its choice, depending on problems of interest. For example, if a multivariate polynomial of $x_i$'s with time-dependent coefficients is chosen as $J(\mathbf{x},t)$, then constancy of $\overline{J}$ gives information on how the moments and correlations between the variables evolve in time. Although the expectation value is conserved, the variance, $(\delta J)^2 = \overline{J^2} - (\overline{J})^2$, depends on time. In fact, from Eqs. (A1) and (A2), we obtain

$$\frac{d(\delta J)^2}{dt} = \frac{d\overline{J^2}}{dt} = 2 \sum_{i,j=1}^{N} \overline{D_{ij} \frac{\partial J}{\partial x_i} \frac{\partial J}{\partial x_j}}. \qquad (A.5)$$

This equation corresponds to Eq. (19) in the quantum-mechanical case and shows that the fluctuation of the classical weak invariant associated with the Fokker-Planck equation also monotonically grows in time. The drift term does not contribute to the growth rate of the fluctuation of the weak invariant, analogously to the fact that the subsystem Hamiltonian does not contribute to Eq. (19), either.

*Note Added*. It has been pointed out by the anonymous referee that the structure in the third expression in Eq. (A.5) appears in discussions about the conditional moment closure methods for dissipation of scalar transport in turbulent flows (see Ref. [52]). This fact may suggest a possible role of a weak invariant in such a research area.